\documentclass{llncs}
\usepackage[utf8]{inputenc}
\usepackage[english]{babel}
\usepackage{parskip}

\usepackage{amsmath}
\usepackage{amssymb}
\usepackage{amsfonts}
\usepackage{stmaryrd}
\usepackage{xcolor}
\usepackage{algorithm}
\usepackage{algpseudocode}
\usepackage{xspace}
\usepackage{mathtools}
\usepackage{enumerate}
\usepackage{hyperref}
\usepackage{bm}

\usepackage{float}
\usepackage{placeins}
\usepackage{caption} 
\usepackage{booktabs}
\usepackage{multirow}

\usepackage{xurl} 
\usepackage{comment}
\newcommand{\Fq}{\mathbb{F}_q}
\newcommand{\Fqm}{\mathbb{F}_{q^m}}

\newcommand{\Fqmn}{\mathbb{F}_q^{m\times n}}

\renewcommand{\vec}[1]{\mathbf{#1}}

\newcommand{\bv}{\vec{b}}

\newcommand{\ev}{\vec{e}}
\newcommand{\gv}{\vec{g}}
\newcommand{\hv}{\vec{h}}

\newcommand{\sv}{\vec{s}}

\newcommand{\vv}{\vec{v}}
\newcommand{\xv}{\vec{x}}

\newcommand{\Av}{\vec{A}}
\newcommand{\Bv}{\vec{B}}
\newcommand{\Cv}{\vec{C}}
\newcommand{\Ev}{\vec{E}}

\newcommand{\Iv}{\vec{I}}

\newcommand{\Mv}{\vec{M}}

\newcommand{\Pv}{\vec{P}}
\newcommand{\Qv}{\vec{Q}}

\newcommand{\Vv}{\vec{V}}

\newcommand{\Xv}{\vec{X}}
\newcommand{\Yv}{\vec{Y}}

\newcommand{\Cvec}{\mathcal C_{vec}}
\newcommand{\Dvec}{\mathcal D_{vec}}
\newcommand{\Amat}{\mathcal A_{mat}}
\newcommand{\Cmat}{\mathcal C_{mat}}
\newcommand{\Dmat}{\mathcal D_{mat}}

\newcommand{\Gmat}{\mathcal G_{mat}}
\newcommand{\Cpub}{\mathcal C_{pub}}

\newcommand{\Cmatd}{\mathcal C_{mat}^\perp}

\newcommand{\A}{\mathcal A}
\newcommand{\B}{\mathcal B}
\newcommand{\C}{\mathcal C}

\newcommand{\E}{\mathcal E}
\newcommand{\G}{\mathcal G}
\newcommand{\K}{\mathcal K}

\newcommand{\R}{\mathcal R}
\newcommand{\Fcal}{\mathcal F}

\newcommand{\BFqm}{\mathcal{B}(\Fqm)}
\newcommand{\oneto}[1]{\{1,..., #1 \}}
\newcommand{\stab}{\text{Stab}}

\newcommand{\norme}[1]{\| #1 \|}
\newcommand{\eqdef}{\stackrel{\textup{def}}{=}}
\newcommand{\floor}[1]{\left\lfloor #1 \right\rfloor}
\DeclareMathOperator{\rank}{rank}
\DeclareMathOperator{\tr}{tr}

\newcommand{\Fold}{\mathsf{Fold}}
\newcommand{\Unfold}{\mathsf{Unfold}}

\newcommand{\MR}{\mathsf{MinRank}}
\newcommand{\MRS}{\mathsf{MinRankSyndrome}}

\newcommand{\addremove}{\mathsf{AddRemove}}
\newcommand{\getsr}{		\mathrel{\vbox{\offinterlineskip\ialign{
				\hfil##\hfil\cr
				\hspace{0.1em}$\scriptscriptstyle\$$\cr
				$\longleftarrow$\cr
			}}}}

\newcommand{\romaric}[1]{\todo[inline,color=green!20]{\textbf{Romaric:} #1}}

\date{ }

\begin{document}

\title{How to break the Miranda signature scheme over matrix Gabidulin codes}
\author{Adrien Vinçotte \thanks{adrien.vincotte@univ-rennes.fr}}
\institute{IRMAR, Université de Rennes, France}
	
\maketitle

\begin{abstract}

The Miranda signature scheme \cite{miranda} relies on masking a matrix code which disposes of a masked underlying structure, and the knowledge of which allows for efficient error decoding. We consider a Gabidulin code (a vector $\Fqm$-linear code) which is expanded into a matrix code, that is only $\Fq$-linear. An additional masking is then applied to it. The attack proposed here shares similarities with that of \cite{Le26} on the EGMC encryption scheme \cite{egmc}, which also follows the paradigm described above. It consists in recovering the $\Fqm$-linear structure of a masked matrix Gabidulin code by reducing to a MinRank instance to be solved over the extension field $\Fqm$, but where the matrices have coefficients in $\Fq$. Such an instance can be efficiently solved. However, unlike the previous attack, it is possible to reduce in polynomial time to such a MinRank instance in the case of the Miranda signature scheme. This results in a particularly efficient key recovery attack against the parameters proposed for Miranda. For example, for the proposed parameter set with $m=79$, the complexity drops from 146 bits to 46 bits in this attack.
    
\end{abstract}
\section{Introduction}

\subsubsection{The Two Notions of Rank Metric Codes.}

The rank metric was initially defined in 1951 by Loo-Keng Hua~\cite{Hua51} as an arithmetic distance on a matrix space with coefficients in $\Fq$. The distance between two matrices $\Av$ and $\Bv$ is defined as the rank of $\Av-\Bv$. In 1978, Delsarte proposed the notion of matrix code~\cite{Del78}~: an $\Fq$-linear subspace of the set of matrices $\Fq^{m\times n}$. The weight of a matrix then corresponds to its rank.

Several years later, Ernst Gabidulin extended the definition of rank metric to the space $\Fqm^n$, enabling the construction of codes that benefit from $\Fqm$-linearity~\cite{Gab85}. The weight of a vector is then the dimension of the $\Fq$-vector space spanned by its coefficients. He then defined Gabidulin codes, which can be viewed as the rank metric analogue of Reed-Solomon codes. The $\Fqm$-linearity of such a code allows for a much more compact definition, thereby mitigating one of the main issues in code-based cryptography: the size of the code representation. He then proposed masking such a code to construct the GPT cryptosystem, an adaptation of the McEliece cryptosystem to the rank metric~\cite{GPT91}. However, the $\Fqm$-linearity introduces a high degree of algebraic rigidity in return, which can be exploited to form particularly effective structural attacks: this is the case for Overbeck's attack~\cite{Ove08}, which efficiently recovers the masked Gabidulin code.

\subsubsection{Two Protocols Based on a Masked $\Fqm$-Linear Matrix Code.}

In order to break the $\Fqm$-linearity structure of a Gabidulin code while retaining its decoding properties, the authors of~\cite{egmc} propose converting the Gabidulin code into an $\Fq$-linear matrix code: they write the vectors of $\Fqm^n$ as matrices in $\Fqmn$, where the columns correspond to the representation of the vector's coordinates in an $\Fq$-base $\gamma$ of $\Fqm$. This matrix code is then masked by adding random coefficients to each element of a basis of rows and columns, which are propagated through left- and right-multiplication by invertible matrices $\Pv$ and $\Qv$ with entries in $\Fq$, yielding an \textit{Enhanced Gabidulin Matrix Code} (EGMC). This family of codes is then used to build the EGMC-McEliece and EGMC-Niederreiter cryptosystems, whose security relies on solving a MinRank instance or recovering the private key from the public key. A first structural attack was proposed by~\cite{pwl}, necessitating a revision of the parameters.

Later, the Miranda signature scheme~\cite{miranda} was designed by also exploiting the transformation of an $\Fqm$-linear Gabidulin code into a matrix code. However, the masking mechanism differs: the authors introduce the $\addremove(\ell_a,\ell_s)$ transformation, which consists in constructing a subcode of codimension $\ell_s$ from the matrix Gabidulin code, and adding $\ell_a$ random matrices to its basis. Miranda is a \textit{hash-and-sign} signature scheme built following the GPV paradigm~\cite{GPV08}: the public code is constructed using the $\addremove(\ell_a,\ell_s)$ transformation with $\ell_a>\ell_s$, implying that it is possible to decode any element of the space, since we operate above the unique decoding regime (though it remains necessary to prove that the signature distribution reveals no information about the underlying Gabidulin code). As in the EGMC encryption scheme, forging a signature requires solving an instance of the MinRank problem: a random syndrome $\sv$ is sampled, and the signature consists of an error $\Ev\in\Fqmn$ associated with $\sv$. This yields a very short signature size, but at the cost of a signing time that is exponential in $\ell_s$ and the decoding capacity of the Gabidulin code.

\subsubsection{The Attack by~\cite{Le26} on the EGMC encryption scheme.}

This recent attack proposes guessing the impact of one of the matrices $\Pv$ or $\Qv$, and computing that of the second using the remaining traces of the $\Fqm$-linearity underlying the public matrix code. This leads to two different variants of the attack depending on whether one guesses $\Pv$ or $\Qv$, both allowing the recovery of an equivalent secret key, which is sufficient to execute the decryption algorithm. In particular, guessing the role of $\Qv$ allows puncturing the public code so as to obtain a dual of dimension 1 in the extension $\Fqm$. It is then possible to set up a bilinear system whose unknowns are the basis $\gamma$ into which the Gabidulin code was expanded, as well as the vector $\hv$ defining the dual. Solving this system yields the masked $\Fqm$-linearity, and hence the secret key. This attack does not break the protocol, but a parameter update proves necessary.

\subsubsection{Contributions.}

This is the first proposed attack against the Miranda signature scheme. We propose here to apply an idea similar to the attack in~\cite{Le26} by puncturing the public code until obtaining a code whose dual has dimension 1 over $\Fqm$. In stark contrast to the EGMC cryptosystem, puncturing is immediate because there is no secret matrix interfering with it. The $\addremove$ transformation leaves enough traces of the underlying $\Fqm$-linearity of the matrix Gabidulin code that one can reduce in polynomial time to solving a MinRank instance that requires finding a rank-1 matrix. Although this instance must be solved over the extension $\Fqm$ (which is large in the given parameters), algebraic attacks based on minor computations are very efficient when looking for a low-rank matrix. This significantly impacts the security of the parameters provided in~\cite{miranda}: for instance, the present attack has a complexity below 100 bits for parameters for which the authors claimed a security level of 143 bits, and drops to 62 bits for those with the smallest value for $m$.
\section{Preliminaries}\label{prelim}

The following notations will be used :

\begin{itemize}
    \item The vectors will be denoted by bold lowercase letters ($\xv \in \Fq^n$), and matrices by bold uppercase letters ($\Xv \in \Fq^{m \times n}$).
    \item We denote by $\Unfold$ the application that consists on writing a matrix of $\Fqmn$ as a vector of $\Fq^{mn}$, by concatenating all the rows of the matrix: \begin{align*}
        \Unfold: \quad \Fqmn &\longrightarrow \Fq^{mn} \\
        \Mv&\longrightarrow (M_{1,1},...,M_{1,n},M_{2,1},...,M_{m,n})
    \end{align*}
    The reverse application is denoted by $\Fold_{m\times n}$.
    \item For $x\in\Fqm$, we denote by $x^{[i]}$ the value of $x$ on which the Frobenius has been applied $i$ times. For a vector $\xv\in\Fqm^n$, $\xv^{[i]}$ does mean that the Frobenius has been applied $i$ times on each coefficient: $$\xv^{[i]} = (x_1^{q^i},...,x_n^{q^i}).$$
    This notation will also be used in the case of matrices.
    \item $\BFqm$ denotes the set of $\Fq$-bases of $\Fqm$.
    \item The dual basis of $\gamma\in\BFqm$ is denoted by $\gamma^*$ (see Definition \ref{trace}).
\end{itemize}

\subsection{Rank metric codes}

\begin{definition}[$\gamma$-expansion]\label{g-e}
  Let $\gamma = (\gamma_1,\dots,\gamma_m)$ be an $\Fq$-basis of $\Fqm$. The $\gamma$-expansion of an element in $\Fqm$ to a vector in $\Fq^m$ is defined as the application:$$M_\gamma: x \in \Fqm \mapsto (x_1,\dots,x_m) \in \Fq^m $$ such that $x = \sum_{i=1}^{m} x_i \gamma_i$.
\end{definition}

$M_\gamma$ extends naturally to a vector $\xv \in \Fqm^n$ and turns it into a matrix $M_\gamma(\xv)\in\Fqmn$, by writing in columns the coordinates of each element in basis $\gamma$ in column. 

\begin{definition}[Vector codes with rank metric]
A vector code $\Cvec$ of parameters $[n,\kappa]_{q^m}$ is an $\Fqm$-subspace of $\Fqm^n$ of dimension $\kappa$ endowed with the rank metric. The weight of a vector $\xv\in\Fqm^n$ is the rank of the matrix $M_\gamma(\xv)$, for a basis $\gamma\in\mathcal{B}(\Fqm)$. The weight of a vector is independent of the choice of the basis $\gamma$.
\end{definition}

Equivalently, the \emph{rank of a vector $\xv$} is the dimension of the $\Fq$-vector space spanned by its coordinates, that is the \emph{support} of $\xv$.
We denote the weight of a vector $\xv$ by:
$$\norme{\xv} \eqdef \rank (M_\gamma(\xv))=\dim (\langle x_1,...,x_n\rangle_q)$$

\begin{definition}[Matrix codes]
A matrix code $\Cmat$ of parameters $[m\times n,k]_q$ is an $\Fq$-vectorial subspace of $\Fqmn$ of dimension $k$ endowed with the rank metric.
\end{definition}

An $\Fqm$-linear vector code $\Cvec$ of parameters $[n,\kappa]_{q^m}$ can be turned into a matrix code defined as: $$M_\gamma(\Cvec) := \{ M_\gamma(\xv) \,:\, \xv \in \Cvec\}.$$

Let $\gamma\in\mathcal{B}(\Fqm)$, and $(\vv_i)_{i\in\oneto{\kappa}}$ be an $\Fqm$-basis of $\Cvec$, then an $\Fq$-basis of $M_{\gamma}(\Cvec)$ is given by: $$\Big\lbrace bM_{\gamma}(\vv_i)~:~ b\in\gamma, i\in\oneto{\kappa}\Big\rbrace.$$
It implies that if $\Cvec$ is a vector code of parameters $[n,\kappa]_{q^m}$, then $M_\gamma(\Cvec)$ is a matrix code of parameters $[m\times n,m\kappa]_q$. In this case, we talk about an "$\Fqm$-linear matrix code". However, it is possible to recover this linearity by calculating the set of matrices that leave the code invariant when multiplying at the left.

\begin{definition}[Left stabilizer]
    Let $\Cmat\subset\Fqmn$ be a matrix code. The left stabilizer of $\Cmat$ is: $$\stab_L(\Cmat)\eqdef\{\Av\in\Fq^{m\times m} : \Av\Cmat\subset\Cmat\}.$$
\end{definition}

While a random code will have a trivial stabilizer with overwhelming probability (and thus of dimension 1: the space spanned by the identity matrix), that of an $\Fqm$-linear code $M_\gamma(\Cvec)$ will contain the matrices of multiplication by elements of $\Fqm$ written in basis $\gamma$ (thus of dimension $m$).

\begin{definition}[Equivalent codes]
    Two vector codes $\Cvec$ and $\Dvec$ are said to be \emph{equivalent} if the exists a matrix $\Qv\in \mathbf{GL}_{n}(\Fq)$ such that $$\Dvec=\Cvec\Qv.$$    
    Two matrix codes $\Cmat$ and $\Dmat$ are said to be \emph{equivalent} if there exist two matrices $\Pv \in \mathbf{GL}_{m}(\Fq)$ and $\Qv \in \mathbf{GL}_{n}(\Fq)$ such that $$\Dmat=\Pv\Cmat\Qv.$$ If $\Pv = \Iv_m$ (resp. $\Qv = \Iv_n$), $\Cmat$ and $\Dmat$ are said to be right equivalent (resp. left equivalent).
\end{definition}

As seen above, the transformation of a vector code into a matrix code is dependent on the choice of the basis $\gamma$. However, two different bases produce left-equivalent codes. For two bases $\beta$ and $\gamma$, if we denote $\Pv$ the transition matrix between $\beta$ and $\gamma$, we get:$$M_\gamma(\Cvec) = \Pv\, M_\beta(\Cvec).$$

On the other hand, the following assertions can be easily checked. For $a\in\Fqm^*$ and $\gamma\in\BFqm$, then $M_{a\gamma}(\Cvec)=M_{\gamma}(\Cvec)$. Also, for $i\in\mathbb{N}$, then $M_{\gamma^{[i]}}(\Cvec)=M_{\gamma}(\Cvec^{[m-i]})$. We consider the action of multiplication by a scalar and the composition by Frobenius on $\BFqm$, and the underlying equivalence relation $\sim$: $$\beta\sim\gamma\iff\text{there exists $a\in\Fqm^*$ and $i\in\mathbb{N}$ s.t. }\gamma=a\beta^{[i]}.$$

We can verify that for a given $\Fqm$-linear matrix code $\Cmat$, then for two basis $\beta,\gamma\in\BFqm$, we have $\beta\sim\gamma$ if and only if the vector codes $M_\beta^{-1}(\Cmat)$ and $M_{\gamma}^{-1}(\Cmat)$ are equivalent.

\subsubsection{Dual of a matrix code.}

If the notion of duality is well-known for a vector code (that is the orthogonal of the canonical scalar product in $\Fqm^n$), it is also necessary to introduce a notion of orthogonality in matrix spaces.

\begin{definition}[Dual of a matrix code]\label{def:dual_matrix_code}
Let $\Cmat$ a matrix code of parameters $[m\times n,k]_q$. Its dual is the matrix code of parameters $[m\times n,mn-k]_q$: $$\Cmatd = \left\lbrace \Yv\in\Fqmn \;:\; \forall\Xv\in\Cmat\; \tr (\Xv\Yv^t)=0 \right\rbrace.$$
\end{definition}

The considered scalar product on matrix spaces does not seem to respect $\Fqm$-linearity, and in general, the dual of an $\Fqm$-linear matrix code loses his linearity on the extension.

\subsection{MinRank problem}

The MinRank problem can be seen as the Rank Decoding problem adapted to $\Fq$-linear matrix codes.

\begin{definition}[MinRank problem]
Given as input matrices $\Yv,\Mv_1,\dots,\Mv_k\in\Fq^{m\times n}$, the $\MR (q,m,n,k,r)$ problem asks to find $x_1,\dots,x_k\in\Fq$ and $\Ev\in\Fq^{m\times n}$ with $\rank\Ev\leq r$ such that $\Yv = \sum_{i=1}^k x_i\Mv_i + \Ev$.
\end{definition}

The decoding problem for a noisy codeword matrix code of parameters $[m\times n,k]_q$ is exactly the $\MR (q,m,n,k,r)$ problem, where $r$ is the rank of the error. Similarly to Niederreiter encryption, the Miranda signature requires decoding a syndrome of the matrix codeword.

We can see that $\tr (\Xv\Yv^t) = \langle\Xv,\Yv\rangle$, where $\langle\cdot,\cdot\rangle$ denotes the canonical scalar product on $\Fqm^n$, allowing one to properly define a notion of dual for matrix codes. The interested reader can find more details in~\cite{miranda}.

\begin{definition}[MinRank-Syndrome problem]
Given as input vectors $\sv,\vv_1$, $\dots,\vv_k\in\Fq^{nm-k}$, the $\MRS (q,m,n,k,r)$ problem asks to find $(e_1,\dots,e_{nm})\in\Fq^{nm}$ with $\rank\Fold _{m\times n}(\ev)\leq r$ such that $\sv = \sum_{i=1}^k e_i\vv_i$.
\end{definition}

The two problems are equivalent for the same reasons that the decoding problem and the syndrome decoding problem are in the vector codes context.

\subsection{Gabidulin codes}

Gabidulin codes were introduced by Ernst Gabidulin in 1985~\cite{Gab85}. These vector codes can be seen as the analog in rank metric of the Reed-Solomon codes, where the codeword is a set of evaluation points of a $q$-polynomial rather than a standard polynomial.

\begin{definition}[$q$-polynomial]
  A $q$-polynomial of $q$-degree $r$ is a polynomial in $\Fqm[X]$ of the form:
  $$P(X) = \sum_{i=0}^{r}p_iX^{[i]} \qquad \text{with } p_r \neq 0.$$
\end{definition}
For a $q$-polynomial $P$, we denote by $\deg_q P$ its $q$-degree.

A $q$-polynomial is also called a \emph{linearized polynomial} since it induces an $\Fq$-linear application due to the linearity of the Frobenius
endomorphism. Due to their structure, $q$-polynomials are inherently related to decoding problems in the rank metric.

\begin{definition}[Gabidulin code \cite{Gab85}]
  Let $\kappa,m,n\in\mathbb{N}$, such that $k\leq n\leq m$. Let
  $\gv=(g_1,\dots,g_n)\in\Fqm^n$ a vector of $\Fq$-linearly
  independent elements of $\Fqm$. The Gabidulin code
  $\mathcal{G}_\kappa (\gv)$ is the vector code of parameters
  $[n,\kappa]_{q^m}$ defined by:
$$\mathcal{G}_\kappa (\gv) = \left\lbrace P(\gv) : \deg_q P < \kappa\right\rbrace,$$ where $P(\gv)=(P(g_1),\dots,P(g_n))$ and $P$ is a $q$-polynomial.
\end{definition}

The vector $\gv$ is said to be an \emph{evaluation vector} of the Gabidulin code $\mathcal{G}_{\kappa} (\gv)$. Gabidulin codes are popular in cryptography because they benefit from a very efficient decoding algorithm achieving the Singleton bound.

 \begin{proposition}[\cite{Gab85}]\label{propo:decoGab}
 	Given a Gabidulin code $\mathcal{G}_\kappa (\gv)$ of parameters $[n,\kappa]_{q^m}$, there exists a deterministic algorithm $\mathsf{Decode}^{\G}$ running in $O(n^{2})$ operations in $\Fqm$ and such that given~$\vec{y} \in \Fqm^{n}$:
 	\begin{itemize}
 		\item if there exists $\vec{c} \in \mathcal{G}_\kappa (\gv)$ and $\vec{e} \in \Fqm^n$ with $\norme{\vec{e}} \leq \frac{n-\kappa}{2}$ such that $\vec{y} = \vec{c} + \vec{e}$, then it outputs $\vec{e}$;		
 		\item otherwise, it outputs $\bot$~. 
 	\end{itemize}
 \end{proposition}

However, their strong structure makes them difficult to hide: Overbeck's attack \cite{Ove08} made it possible to break a very large number of Gabidulin code maskings. On the other hand, while the $\Fqm$-linearity of a matrix code is lost when passing to the dual, this is not the case for a Gabidulin code: the dual of $M_\gamma(\G)$ is indeed an other matrix Gabidulin code, which corresponds to the code $\G^\perp$ expanded in the dual base of $\gamma$.

\begin{definition}[Trace application and dual basis]\label{trace}
    Let Trace: $\Fqm\rightarrow\Fq$ be the $\Fq$-linear application given by:
    $$ \tr: x\longrightarrow  \quad\sum_{i=0}^{m-1}x^{[i]}.$$
The dual basis of $\gamma = (\gamma_1,...,\gamma_m)\in\BFqm$ is the only basis $\gamma^* = (\gamma^*_1,...,\gamma^*_m)\in\BFqm$ such that $\tr(\gamma_i\gamma^*_j)=\delta_{i,j}$ for all $j\in\oneto{m}$.
\end{definition}

We usually denote by $\gamma^*$ the dual basis of $\gamma\in\BFqm$. The following properties can be easily check. For $\gamma\in\BFqm$ and $a\in\Fqm^*$, then $(a\gamma)^*=a^{-1}\gamma^*$. Also, for $i\in\mathbb{N}: \left(\gamma^{[i]}\right)^* = \left(\gamma^*\right)^{[i]}$. It implies that for $\beta,\gamma\in\BFqm$: $$\beta\sim\gamma\quad\iff\quad\beta^*\sim\gamma^*.$$

\begin{proposition}[Dual of a matrix Gabidulin code~\cite{Rav16}]
    Let $\G$ be a Gabidulin code, $\gamma\in\BFqm$. Then: $$M_\gamma(\G)^\perp = M_{\gamma^*}(\G^\perp).$$
\end{proposition}

This implies that passing to the dual of a matrix Gabidulin code will leave traces of $\Fqm$-linearity. Also, the link between equivalent bases for the relation $\sim$ and equivalent matrix codes mentioned above is compatible by passing to the dual in the case of a Gabidulin code. Let $\Gmat$ be a matrix Gabidulin code (extended in any basis of $\Fqm$) and two bases $\beta,\gamma\in\BFqm$. Then we have $\beta\sim\gamma$ if and only if the vector codes $M_\beta(\Gmat)^\perp$ and $M_\gamma(\Gmat)^\perp$ are equivalent.

\subsection{The Miranda signature scheme}


We present here the construction of the secret key of the Miranda signature scheme. First, a secret Gabidulin code $\G$ of parameters $[n,\kappa]_{q^m}$ is expanded into a $[n \times m,\kappa m]_{q}$ matrix code by applying the transform $M_\gamma$, where $\gamma$ is uniformly sampled from $\BFqm$. Then, the authors of Miranda introduce the $\mathsf{AddRemove}$ transform to mask a secret matrix Gabidulin code, allowing them to break the hidden $\Fqm$-linearity of the matrix code, and prevent the Overbeck attack~\cite{Ove08}.

\begin{definition}[Add-and-Remove matrix-code construction~\cite{miranda}]\label{def:addRemove} Let
  $m, n, k,$ $q,\ell_a, \ell_s$ be integers and~$\Cmat$ be a $[m \times n,\kappa m]_{q}$-code. The \emph{Add-And-Remove} transform $\mathsf{AddRemove}\left( \Cmat,\ell_a,\ell_s \right)$ is defined as follows. Let $\mathcal{C}_{s}$ be an arbitrary subcode of $\Cmat$ of codimension $\ell_{s}$. Then, let~$\Amat$ be an arbitrary $[m \times n,\ell_a]_{q}$-code such that $\mathcal{C} \cap \Amat = \{\vec{0}_{m \times n}\}$. We define the resulting \emph{Add-and-Remove} code $\Dmat$ as follows:
  $$\Dmat\eqdef \mathcal{C}_{s} \oplus \Amat $$ where $\oplus$ denotes the direct sum of vector spaces.
It defines an $[m \times n, \kappa m-\ell_s+\ell_a]_{q}$ matrix code. 
\end{definition}

Miranda public keys are defined as a random basis $(\vec{B}_{1},\dots, \vec{B}_{(n-\kappa)m + \ell_{s}- \ell_{a}})$ of the dual of some code $\Dmat$ obtained via the $\mathsf{AddRemove}$ construction. The key generation algorithm is formally defined in Appendix \ref{keygen}.

Given the hash of the message to sign: $\vec{s} \in \Fq^{(n-\kappa)m+\ell_s-\ell_a}$, a signature is then the matrix $\vec{E} \in\Fqmn$ with rank at most $\floor{\frac{n-k}{2}}$ such that:
$$\left( \tr \left( \vec{E}\vec{B}_{i}^{\top} \right) \right)_{i=1}^{(n-\kappa)m+\ell_s-\ell_a} = \vec{s} \ . $$

The associated trapdoor is the knowledge of the underlying code $\mathcal{C}=M_\gamma(\G)$ to form~$\Dmat$. How to use this trapdoor to compute a valid matrix $\Ev$ is explained in Appendix \ref{sign}.

The authors of Miranda take two attacks into account when calculating the parameters of the signature. The first is the forgery signature, which consists in directly trying to solve the MinRank instance to compute a valid error $\Ev$. On the other hand, the authors of Miranda propose a structural attack which consists in recovering the hidden $\Fqm$-linearity by retrieving the left–stabilizer algebra of the secret code. 

Our goal in the following will be to improve on the current structural attack by leveraging Le's attack~\cite{Le26}. We will see that we manage to recover the $\Fqm$-linearity, which is supposedly hidden, as the codes $\C_s$ and $\Amat$ are only linear over $\Fq$, and not over the extension.
\section{Key recovery attack for Miranda} \label{attaque}

This section introduces another structural attack different from the one previously proposed by the authors of Miranda. It consists of reducing in polynomial time to the resolution of a MinRank instance whose target matrix has rank 1, in a similar fashion to Le's attack~\cite{Le26}.


Let $\G$ be a (secret) Gabidulin code of parameters $[n,\kappa]_{q^m}$, and $\gamma\in\BFqm$. We consider the $[m \times m, \kappa m-\ell_s+\ell_a]$ public code  $$\Cmat = \C_s\oplus\Amat$$ where $\C_s$ is a subcode of $M_\gamma(\G)$ of co-dimension $\ell_s$ and $\Amat$ has dimension $\ell_a$ with $M_\gamma(\G)\cap\Amat = \lbrace 0\rbrace$. We further assume that $\Fqm(\Amat)$ is a $\Fqm$-vector space of dimension $\ell_a$, which occurs with high probability (if it is not the case, the attack can be performed by decreasing $\ell_a$ by 1, until reaching the dimension of $\Amat$). In what follows, we suppose that $\ell_a<n-\kappa$ (at the end of Subsection~\ref{reduc_minrank} we show that we can always reduce ourselves to this case).

The attack consists of truncating the matrix code so that its $\Fqm$-linear vector version has a dual of dimension 1 in $\Fqm$. However, some elements of the attack change due to the nature of the masking. In fact, as the maskings in the ACDGV encryption scheme and Miranda are very different, it is not straightforward that one can use the same methods to attack them. Furthermore, the addition of random matrices in Miranda's public code leads to a change in the behavior of the MinRank solver. 

The implementation in sage of functions allowing to recover the normalized basis $\gamma$ can be found here: \url{https://github.com/AdrienVincotte/Stickelberger-attack-on-the-Miranda-signature-scheme.git}. Recover the secret Gabidulin code from this basis is a well-known problem (see Subsection \ref{recover_code}).

\subsection{Reduction in polynomial time to an easy MinRank instance}\label{reduc_minrank}

The goal of this section is to show that recovering the $\Fqm$ basis (up to multiplication by a scalar) used in the secret key can be reduced to solving a rank $1$ MinRank instance, in a very large field.

The first step of the attack is to sample a random full-rank matrix $\Vv\in\Fq^{m\times (\kappa+\ell_a+1)}$, and then to consider the $[\kappa+\ell_a+1,\kappa]_{q^m}$ Gabidulin code $\G'\eqdef\G\Vv$, which has for evaluation vector $\gv\Vv$. We then define
$$\Cmat' \eqdef\Cmat\Vv$$
 which is a matrix code with parameters $[m\times (\kappa+\ell_a+1),\kappa m-\ell_s+\ell_a]_q$. We also define $\Amat' \eqdef\Amat\Vv$. Then, the code $$M_\gamma^{-1}(\Cmat')=\G' \oplus \Mv_{\gamma}^{-1}(\Amat')$$ spans a $[\kappa+\ell_a+1,\kappa+\ell_a]_{q^m}$ code over $\Fqm^n$, then with a one-dimensional dual. The aim of this attack is to determine a vector $\hv\in\Fqm^{\kappa+\ell_a+1}$ that generates this dual.

Let $(\Cv^{(1)},...,\Cv^{(\kappa m-\ell_s+\ell_a)})$ be a basis of $\Cmat'$. By definition of $\hv$, for all $1\leq s\leq \kappa m-\ell_s+\ell_a$, we have: $$\langle M_\gamma^{-1}(\Cv^{(s)}),\hv\rangle =0.$$

 We deduce that for all $1\leq s\leq \kappa m-\ell_s+\ell_a$: $$\gamma\Cv^{(s)}\hv^T=0.$$

\begin{remark}
    It should be noted that the solution is not unique (possible bases $\gamma$ can belong to different orbits of the relation $\sim$), although the subcode $\C_s$ of the $\Fqm$-linear matrix code does not break underlying $\Fqm$-linearity.
\end{remark}

Expressing this computation in terms of the components of the unknowns $\gamma \in \Fqm^m$ and $\hv \in \Fqm^{\kappa+\ell_a+1}$, we obtain: $$\sum_{i=1}^m\sum_{j=1}^{\kappa+\ell_a+1}\gamma_i h_jc^{(s)}_{i,j}=0\qquad\text{for all }1\leq s\leq \kappa m-\ell_s+\ell_a.$$

Setting $w_{i,j}=\gamma_i h_j\in\Fqm$, we obtain a linear system of $\kappa m-\ell_s+\ell_a$ equations in $\underbrace{m}_{\gamma}(\underbrace{\kappa+\ell_a+1}_{\vec{h}})$ unknowns. This makes it an underdetermined linear system.

We write the system as a matrix $\Av\in\Fq^{(m\kappa-\ell_s+\ell_a)\times m(\kappa+\ell_a+1)}$ (notice that the coefficient of $\Av$ are in $\Fq$). Solving this system is equivalent to finding a vector $\vec{w} \in \Fqm^{\kappa+\ell_a+1}$ such that: $$\vec{w}\in\ker_{q^m}\Av\quad\text{  and  }\quad\rank\Fold_{m\times (\kappa+\ell_a+1)}(\vec{w})=1.$$
Indeed, because of the constraints on the values $w_{i,j}$, $\Fold_{m\times (\kappa+\ell_a+1)}=\gamma^T\hv$, the folded version of $\vec{w}$ has rank 1.

Let $\rho = \dim\ker_{q^m}\Av$. Since all the $\kappa m-\ell_s+\ell_a$ linear equations are independent~\footnote{This is an immediate consequence of the fact that the matrices $(\Cv^{(i)})_i$ are also independent}, the rank theorem implies that $\rho = (m-1)(\ell_a+1)+\ell_s+1$. Let $(\bv_1,...,\bv_\rho) \in \Fqm^{m(\kappa+\ell_a+1)}$ be a basis of $\ker_{q^m}\Av$. For $1\leq z\leq\rho$, we define a basis of the folded kernel: $\Bv_z=\Fold_{(\kappa+\ell_a+1)\times m}(\bv_z)$.

We seek an element of $\ker_{q^m}\Av$ whose folded form is a rank-1 matrix, which corresponds to searching for such a matrix in the matrix code $\langle \Bv_1,...,\Bv_\rho\rangle_{q^m}\subset\Fqm^{m\times \kappa+\ell_a+1}$. We have thus reduced the problem in polynomial time to a MinRank instance in a code with parameters $[m\times(\kappa+\ell_a+1),\rho]_{q^m}$, where the target is a rank-1 matrix.

However, it should be noted that this MinRank instance is solved in a matrix code whose coefficients of a basis belong to $\Fq$, and not $\Fqm$. Then, the set of solutions is stable under multiplication by an element of $\Fqm^*$, and composition by Frobenius. As a result, the MinRank instance is not \emph{random} and the usual solvers are not necessarily accurate. The topic of Subsection~\ref{solveminrank} is thus to explain how we solve the MinRank instance, in the same fashion as Le~\cite{Le26}, but with small subtleties that require to be careful.



\subsubsection{The case $\ell_a = n-\kappa$.}

If the code $M_\gamma^{-1}(\Cpub)$ were to span the entire space $\Fqm^m$, this attack would become impracticable. This corresponds to the case $$\ell_a=n-\kappa=2t.$$However, one can always draw a subcode $\Tilde{\C}$ of $\Cpub$ of codimension $\ell$ until the attack can be applied: it would simply suffice for the intersection between $\Tilde{\C}$ and $\A$ to be of dimension $\ell_a-1$ so that $M_\gamma^{-1}(\Tilde{\C})$ no longer spans the entire space $\Fqm^m$, and then apply the attack to $\Tilde{\C}$. This subcode can be chosen with a small dimension, the only constraint being to preserve the $\Fqm$-linear structure of the code $M_\gamma(\G)$, in order to ensure that its dual has dimension 1 over $\Fqm$.

\begin{remark}
    It is possible that the intersection between $\Tilde{\C}$ and $\A$ has dimension lower than $\ell_a-1$. If it happens, all vectors belonging to a $\Fqm$-vectorial space of dimension 2 will define a solution, and then the kernel of the Macaulay matrix defined in Subsection~\ref{solveminrank} will have an abnormally large dimension. The attacker then simply has to truncate the code to obtain a length equal to $\kappa+\ell_a$ (or eventually smaller if needed) rather than $\kappa+\ell_a+1$ to obtain a dual of dimension~1.
\end{remark}

\subsection{Solving the MinRank instance}\label{solveminrank}

Now that we have shown that one should aim at recovering a rank-1 matrix, we detail here \emph{how} to proceed, that is to say, \emph{how} to solve a MinRank instance, in the matrix code $\K =\langle \Bv_1,...,\Bv_\rho\rangle_{q^m}$ with parameters $[m\times (\kappa+\ell_a+1),\rho]_{q^m}$. The answer seems simple, as it is possible to solve this MinRank instance using existing generic algorithms, such as Minors or SupportMinors (and these would be sufficient to attack the parameters proposed in~\cite{miranda}). However, as highlighted previously, this instance is not generic, since all matrices have coefficients in $\Fq$, and thus the behavior of the generic solvers becomes harder to predict. Nevertheless, one can solve this system very efficiently by performing linear algebra over $\Fq$ rather than over the extension. This part of the attack proceeds almost identically to that of~\cite{Le26}. However, unlike Le's attack, parasitic solutions appear. We will see that this requires additional care in the analysis.

\subsubsection{Step 1: modeling using minors.}

We define the multivariate polynomial ring $A:=\Fqm[x_1,...,x_\rho]$. We write $$\Ev = \sum_{i=1}^\rho x_i\Bv_i\in\K$$ for the targeted rank-1 matrix. By expressing the $2\times 2$ minors of $\Ev$ as functions of the variables $x_i$, we derive a homogeneous quadratic system of equations, consisting of $N=\binom{m}{2}\binom{\kappa+\ell_a+1}{2}$ equations and $\binom{\rho+1}{2}$ monomials, that fully characterize the fact that $\Ev$ is of rank $1$.


We denote by $\Fcal := (f_i)_{i\in\oneto{N}}$ the set of these equations, and by $I:=\langle(f_i)_{i\in\oneto{N}}\rangle$ the ideal they generate in $A$.

Since the matrices $\Bv_i$ have coefficients in $\Fqm$, the set of solutions is stable under Frobenius composition and multiplication by an element of $\Fqm^*$. The set of solutions can therefore be partitioned into $K$ distinct orbits under this action.

Since the extension of a matrix code is identical in bases $\gamma$ and $a\gamma$, where $\gamma\in\BFqm$ and $a\in\Fqm^*$, the set of solutions we are seeking is a subset of one of these orbits. However, Le's attack heavily relies on the conjecture that $\dim_{\Fq}(\ker(\Av)) = m$. Here, this is not the case: the solution orbit is not unique (this is where it differs from the attack of~\cite{Le26}). Still, we manage to make the following adapted conjecture, which will not prevent us from using the same techniques.

\begin{conjecture} \label{conjecture:nbsol}
    If $N > \binom{\rho+1}{2}$, then there are enough non-redundant equations for the system (seen as a linear system) to be overdetermined. It may be that for some parameters, $N< \binom{\rho+1}{2}$. In that case, we consider $\Mv_{d}(\Fcal)$ the Macaulay matrix in degree $d$ of the system, for $d \ge 3$.
\end{conjecture}

In practice, we observe that degree $2$ is sufficient for most of the parameter sets, and degree $3$ is sufficient for all attacked sets of parameters, that is, one only multiplies the polynomials of $\Fcal$ by each variable $x_i$.

With that conjecture in mind, we continue following the steps of Le for the solving of the system. However, the different orbits that are also solutions of the MinRank instance will require an additional cost.

\begin{remark}
    It has been observed that the set of solutions is stable under composition with the Frobenius map. However, this does not imply that the number of solutions in the projective space $\mathbb{P}^{\rho-1}(\Fqm)$ is a multiple of $m$: it may happen to find a solution whose all coordinates belong to the base field $\Fq$ (or more generally in any subfield of $\Fqm$, but $m$ has been chosen prime in the Miranda parameters). However, these solutions will be immediately rejected because the vector $\gamma$ extracted from them cannot be an element of $\BFqm$. This is why we now assume that the number of solutions is a multiple of $m$, those that are incompatible will be simply ignored.
\end{remark}

\subsubsection{Step 2: solving the quadratic system.}

The ideal $I$ under consideration consists of homogeneous equations, and is therefore stable under multiplication by an element of $\Fqm^*$. $I$ is then a union of $\Fqm$-linear lines. We specialize the variable $x_1$ to 1, and define the ideal: $I'=I+\langle x_1-1\rangle$, as well as the algebra $$\A' = \Fqm[x_1,...,x_\rho]/I'.$$
We make the reasonable assumption that it is a discrete variety.

\begin{remark}
    It implies that the attack fails if the associated coefficient to $\Bv_1$ in the solution of the MinRank instance is 0, which happens with a probability of $1/q^m$. This probability is bigger than $2^{-\lambda}$ (where $\lambda$ denotes the security parameter) for the majority of sets of parameters attacked in Section~\ref{application}. Nevertheless, the attack can be done again by specializing another variable to 1 in case of failure.
\end{remark}

We denote the dimension of this algebra by $\dim\A' = \delta$, which corresponds to its cardinality. We then compute a basis $\B = (1,b_2,...,b_{\delta})$ of the algebra $\A'$ using Gaussian elimination on the Macaulay matrix $\Mv\in\Fq^{N\times\binom{\rho+1}{2}}$ associated with the system of equations. This step has a complexity of: $$O\left(\binom{\rho+1}{2}^\omega\right).$$

Note that the coefficients of the quadratic equations belong to $\Fq$ and not to $\Fqm$, all the computations of this step are done in the base field. Performing these operations within the extension would defeat the purpose of this method.

\begin{remark}
    The complexity increases to $O\left(\binom{\rho-1+d}{d}^\omega\right)$ for parameter sets that requires us to consider $\Mv_d(\Fcal)$, with $d\geq 3$, instead of $\Mv(\Fcal)$.
\end{remark}

Specializing this value to obtain a discrete variety (or equivalently $I$ is a 0-dimensional ideal) allows the use of the Eigenvalue Theorem, which will enable the determination of the entire variety using linear algebra operations performed over the base field $\Fq$.

\begin{theorem}[Eigenvalue theorem~\cite{clo98}]\label{eigenth}
    Let $\mathbb{K}$ be an arbitrary field, and $I$ a 0-dimensional ideal of $\mathbb{K}[X_1,...,X_n]$. Then the quotient algebra $$\A_I = \mathbb{K}[X_1,...,X_n]/I$$ is a finite-dimensional $\mathbb{K}$-vector space of dimension $\delta$. Let $(\zeta_1,...,\zeta_\delta)$ denote the set of points of the affine variety defined by $I$.

For $P\in\mathbb{K}[X_1,...,X_n]$, we define the endomorphism of multiplication by $P$: \begin{align*}
    m_P : \A_I &\longrightarrow\A_I \\
    \overline{f}&\longrightarrow\overline{Pf}.
\end{align*}
The $\delta$ eigenvalues of $m_P$ in $\Bar{\mathbb{K}}$ are $(P(\zeta_j))_{j\in\oneto{\delta}}$.
\end{theorem}



Let fix $i\in\oneto{\rho}$, and be $(\lambda^{(i)}_j)_{j\in\oneto{\delta}}$ the set of eigenvalues of $m_{x_i}$. We assume that all eigenvalues are simple. In this case, an eigenvector $\vv_j$ associated to $\lambda^{(i)}_j$ consists of the evaluations of all elements of the basis $\B=(b_1,b_2,...,b_\delta)$ at the point~$\zeta_j$~\cite{clo98}:
$$\vv_j = \left(b_1(\zeta_j),b_2(\zeta_j),...,b_\delta(\zeta_j)\right).$$

It implies that all the applications $(m_{x_i})_i$ (and then all their linear combinations) share the same eigenvectors. This is why this step is independent of the chosen variable $x_i$. Thus, the next step consists in computing these eigenvalues to deduce the set of points $(\zeta_1,...,\zeta_\delta)$. Rather than diagonalize an application $m_{x_i}$ randomly chosen among all the possible, we compute a random linear combination of these applications, that has a greater probability to have simple eigenvalues.

Since $\B$ spans all the quotient, the variables $x_i$ can be expressed in terms of $b_i$ according the coefficients of a matrix $\Cv\in\Fq^{\delta\times\rho}$: $$(x_1,...,x_{\rho}) = \B\Cv.$$

Computing $\vv_i\Cv$ allows one to deduce the solution belonging to the ideal $I'_i$. Repeating the same procedure for the other eigenvalues allows the other possible solutions to be recovered.

\begin{remark}
    If the coefficients of a computed $\vv_j$ all belongs to a strict subfield of $\Fqm$ (in the case of Miranda where $m$ is always prime, that is equivalent to say that $\vv_j\in\Fq^\delta)$), it can be immediately deduced that this is an unwanted solution, and it can therefore be rejected.
    For a valid eigenvector $\vv$, then all its Frobenius conjugates $(\vv^{[l]})_{l\in\oneto{m-1}}$ are also eigenvectors. By denoting $K$ as the number of full orbits in the set of solutions, it implies that this process has to be done only $K$ times, plus the number of unwanted solutions, rather than $\rho$.
\end{remark}


\subsubsection{Step 3: extract a possible basis.}

From each of the $Km$ possible solutions, we recover the unfolded version of the matrix $\Ev=\gamma^T\hv$, from which we easily extract a vector $\gamma$, defined up to a multiplicative factor.

\subsubsection{About the number of unwanted solutions.}

We obtain $Km$ possible solutions for the values of $(x_i)_{i\in\oneto{\rho}}$ such that $\Ev\in\K$ (one orbit per $\lambda_i$ considered previously). Since the Macaulay matrix $\Mv$ has size $N\times\binom{\rho+1}{2}$, we deduce by the rank theorem that the dimension of its kernel is upper-bounded by (there is \emph{at most} $N$ linearly independent equations):
$$\dim\ker\Mv\leq N-\binom{\rho+1}{2} = O(m^2(\kappa+\ell_a)^2)$$
implying that the number of solutions (in the projective space $\mathbb{P}^{\rho-1}(\Fqm)$) is roughly bounded by $m^2(\kappa+\ell_a)^2$.

However, this is a very large upper bound: such a high number of parasitic solutions is never reached in practice. However, it is sufficient to show that the final phase of the attack has negligible complexity compared to the execution of Gaussian elimination on the Macaulay matrix. Table~\ref{exp_kernel} shows some experimental results about the dimension of its kernel. For a given parameter set, it seems to be independent of the public key. Then, when $m$ has been fixed, the dimension $\kappa$ of the code and the codimension $\ell_s$ do not seem to affect the dimension of the kernel.

\begin{table}[ht]
	\begin{center}
		{\setlength{\tabcolsep}{0.3em}
			{\renewcommand{\arraystretch}{1.6}
				{\scriptsize
					\begin{tabular}{|c||c|c|c|c|c|c|c|c|}
						\hline
						$\ell_a$ & 0 & 1 & 2 & 3 & 4 & 5 & 6 & 7 \\ \hline
                        dim ker $\Mv$ & 17 & 34 & 69 & 139 & 261 & 452 & 729 & 1126 \\ \hline
					\end{tabular}
					\vspace{0.2\baselineskip}
		}}}
	\end{center}
	\caption{Experimental computation of dimension of kernel of the Macaulay matrix, for $m=n=17$, $\kappa=9$, $\ell_s=10$, with $\ell_a\in\oneto{n-\kappa+1}$}
	\label{exp_kernel}
\end{table}

\subsection{Recovering the secret masked code}\label{recover_code}

The set of solutions to the MinRank instance then yields a list of possible candidates for the secret basis $\gamma$ (recall that the set of solutions is defined up to multiplication by an element of $\Fqm^*$). The final step of the algorithm consists in guessing this basis, and once the $\Fqm$-linearity has been recovered, the secret Gabidulin can be computed.

The public key of the Miranda signature scheme consists of the dual of a matrix code of the form $\addremove (M_\gamma(\G),\ell_a,\ell_s)$. Since $M_\gamma(\G)^\perp = M_{\gamma^*}(\G^\perp)$, this means that the public key is a code of the form $\addremove (M_{\gamma^*}(\G^\perp),\ell_s,\ell_a)$. Consequently, we compute the dual basis of all possible candidates (recall that the dual of a basis is compatible with multiplication by an element of $\Fqm^*$).

Assume we guess the basis $\gamma$. It is possible to compute the dual basis, and then the left-stabilizer algebra $\A$ of the secret matrix code $M_\gamma(\G)^\perp$. We use it to compute the $\Fqm$ space spanned by $\Cmat^\perp$. We have: $$\Cmat^\perp = \E \oplus \R_s$$ where $\E$ is a subcode of $M_{\gamma}(\G)^\perp$ and $\R_s$ is a random matrix code.

We then compute: $$\A\Cmat^\perp = \A\E \oplus \A\R_s$$ where $\A\E$ will be with overwhelming probability the full dual $M_{\gamma}(\G)^\perp$ and $\A\R_s$ will be a random $\Fqm$-linear matrix code.

Let $\Dvec = M_\gamma^{-1}(\A\Cmat^\perp)$. Recall that we do not have the basis $\gamma$, but only an equivalent basis, which implies that this computation leads to several possible codes, but that are all equivalent. By computing the dual of $\Dvec$, we obtain a subcode of the secret Gabidulin code $\G$. Then, the corresponding Gabidulin code can be recovered from the subcode using Overbeck–like techniques in polynomial time~\cite{Ove08}.

\subsection{Complexity of the attack}

All computation required by this attack can be done in polynomial time. The main computation relies on the gaussian elimination on the Macaulay matrix $\Mv$ of the system, that requires $O\left(\binom{\rho-1+d}{d}^\omega\right)$ operations in $\Fq$, for $d \in \{2,3\}$ (depending on the parameter sets). The computation of the kernel of the matrix $\Av$ (see Subsection~\ref{reduc_minrank}) and the Overbeck-like attack to recover the original Gabidulin code $\G$ (although it is necessary to do the last computation $\rho$ times) are negligible compared to that of simplifying the ideal. Then, the complexity is: $$O\left(\binom{(m-1)(\ell_a+1)+\ell_s+d}{d}^\omega\right)$$ where $\omega$ denotes the complexity algebra constant and where $d$ is the smallest value such that the number of equations obtained is greater than the number of unknowns.
\section{Concrete application to Miranda Parameters}\label{application}

\subsubsection*{Signature scheme.}

In the case of the Miranda signature scheme, the public key is the dual of a matrix code of the form $\addremove (M_\gamma(\G),\ell_a,\ell_s)$. Since $M_\gamma(\G)^\perp = M_{\gamma^*}(\G^\perp)$, this means that the public key is a code of the form $\addremove (M_{\gamma^*}(\G^\perp),\ell_s,\ell_a)$. One can therefore directly run the attack on this code, the difference is that the recovered basis is the dual basis of the secret one.

It results a bilinear system of $\binom{m}{2}\binom{2t+\ell_s+1}{2}$ equations and $\rho$ unknowns (and then $\binom{\rho}{2}$ distinct monomials after specialization of $x_1$ to 1), where $\rho=(m-1)(\ell_s+1)+\ell_a+1$. We can then apply the attack detailed in Section~\ref{attaque}. The resulting complexities are given in Table~\ref{comp_mir}. The Forge attack consists on solving the MinRank instance without trying to recover the underlying structure of the matrix code. The "Struct." abbreviation relies on the structural attack proposed by the authors of Miranda.

\begin{table}[ht]
\begin{center}
{\setlength{\tabcolsep}{0.3em}
    {\renewcommand{\arraystretch}{1.6}
        {\scriptsize
            \begin{tabular}{|c|c|c|c|c||c|c||c|c|}
                \hline
                $m$ & $\kappa$ & $\;\; t\;\;$ & $\;\ell_a\;$ & $\;\ell_s\;$& Forge. & Struc. & $\;\; d\;\;$ & This\\ \hline\hline
                79 & 71 & 4 & 189 & 2 & 146 & 146 & 2 & 46 \\ \hline
                113 & 107 & 3 & 220 & 3 & 145 & 143 & 2 & 49 \\ \hline
                127 & 121 & 3 & 250 & 2 & 156 & 160 & 2 & 49 \\ \hline
                223 & 219 & 2 & 330 & 4 & 149 & 147 & 3 & 81 \\\hline\hline
                127 & 119 & 4 & 315 & 2 & 221 & 217 & 2 & 50 \\ \hline
                173 & 167 & 3 & 340 & 4  & 207 & 207 & 2 & 54 \\ \hline
                179 & 173 & 3 & 350 & 2 & 210 & 212 & 2 & 52 \\ \hline\hline   
                163 & 155 & 4 & 400 & 2 & 277 & 276 & 2 & 52 \\ \hline
                233 & 227 & 3 & 160 & 5 & 270 & 270 & 2 & 57 \\ \hline
                239 & 233 & 3 & 470 & 4 & 277 & 273 & 2 & 56 \\ \hline
            \end{tabular}
            \vspace{0.2\baselineskip}
}}}
\end{center}
	\caption{Complexity of parameters given by~\cite{miranda} for Miranda taking  $\omega = 2.8$ as the complexity exponent of linear algebra.}
	\label{comp_mir}
\end{table}

\subsubsection*{Encryption scheme.}

The authors of Miranda also propose a Niederreiter-like encryption scheme relying on the same masking: they keep secret a matrix Gabidulin code on which they apply the $\addremove$ transform. The idea is to consider a decoding distance such that there is uniqueness of the solution, that implies to choose a big value for $\ell_s$ and a small one for $\ell_a$. Furthermore, the authors of Miranda suggest choosing $\ell_a=0$, since the decoding process requires performing an exhaustive search over a random code $\mathcal{A}$ of size $q^{\ell_a}$. It implies that the public code is a subcode of the secret matrix Gabidulin code $M_\gamma(\G)$, and the public key consists of the dual of this matrix code. The message to encrypt is hashed into a matrix of rank at most $t$ (which corresponds to the decoding capacity of $M_\gamma(\G)$), and the ciphertext is its associated syndrome.

As for the signature scheme, the public key is a code of the form $\addremove (M_{\gamma^*}(\G^\perp),0,\ell_s)$. Nevertheless, we will attack the original code $\addremove (M_{\gamma}(\G),\ell_s,0)$ rather than its dual given in the public key: the $\Fqm$-space that it spans has a smaller dimension, allowing to do a bigger truncation to the code, and then implying a MinRank instance with smaller parameters to solve. The resulting complexities are given in Table~\ref{comp_enc}.

\begin{table}[ht]
	\begin{center}
		{\setlength{\tabcolsep}{0.3em}
			{\renewcommand{\arraystretch}{1.6}
				{\scriptsize
					\begin{tabular}{|c|c|c|c||c|c||c|c|}
						\hline
						$m$ & $\kappa$ & $\; t\;$ & $\;\ell_s\;$ & Forge & Struc. & $\;\; d\;\;$ & Comp.\\ \hline\hline
						37 & 25 & 6 & 171 & 141 & 143 & 2 & 68  \\ \hline
						41 & 31 & 5 & 159 & 158 & 143 & 2 & 68  \\ \hline
						43 & 35 & 4 & 164 & 146 & 160 & 2 & 69  \\ \hline
						53 & 47 & 3 & 188 & 153 & 193 & 2 & 71  \\ \hline\hline
						53 & 43 & 5 & 215 & 206 & 208 & 2 & 72  \\ \hline
						59 & 51 & 4 & 208 & 214 & 207 & 2 & 73  \\ \hline
						71 & 65 & 3 & 195 & 211 & 208 & 2 & 74  \\ \hline\hline
						67 & 57 & 5 & 320 & 288 & 272 & 2 & 74 \\ \hline
						97 & 91 & 3 & 360 & 288 & 272 & 2 & 82 \\ \hline
					\end{tabular}
					\vspace{0.2\baselineskip}
		}}}
	\end{center}
	\caption{Complexity of parameters given by~\cite{miranda} for encryption scheme, $\ell_a=0$, $\omega=2.8$.}
	\label{comp_enc}
\end{table}
\section*{Acknowledgement}

The author thanks Romaric Neveu for helpful conversation and for its careful reviewing of the paper. This work is supported by Agence Nationale de la Recherche through grant number ANR-24-CPJ1-0075-01, and by the DGA via CreachLab.

\newpage

\bibliographystyle{alpha}
\bibliography{biblio}

\newpage

\appendix
\section{Key generation of the Miranda signature scheme \cite{miranda}}\label{keygen}

Here is presented the formal key generation algorithm of the Miranda signature scheme. It takes in input the parameters of the scheme. It outputs the public code and the trapdoor, which consists on the secret information (the structure of the $\Fqm$-linearity and the masked Gabidulin code) allowing to compute the error associated to a syndrome for the public code.

\begin{algorithm}[htb]
 	\caption{$\mathsf{KeyGen}\left(\lambda\right)$: Miranda key generation algorithm outputting the public code $\vec{B}_1,\dots,\vec{B}_{mn-km+\ell_{s}-\ell_{a}}$ and its corresponding trapdoor $T$}\label{algo:trap}\medskip 
 	Parameters: $q,m,n,\kappa, \ell_{a},\ell_{s}$ to ensure $\lambda$ bits of security\medskip
 	\hrule
 \setlength{\baselineskip}{1.25\baselineskip}
 	\begin{algorithmic}[1]
 		\State $ \vec{g} \getsr \left\{ (x_1,\dots,x_n) \in \Fqm^{n}\mbox{ whose entries are $\Fq$--linearly independent} \right\}$ 
 		\State $\gamma \getsr \BFqm$
 		\State Compute $\mathcal{C} \eqdef M_{\gamma}(\G_\kappa(\gv))$
 		\State\label{state:Cs} $\mathcal{C}_{s} \getsr \left\{\text{subcodes of } \mathcal C \text{ of codimension } \ell_{s}\right\}$
 		\State\label{state:Amat} $\Amat \getsr \left\{ \mathcal{U}: \mbox{ $[m\times n,\ell_{a}]_{q}$-codes such that $\mathcal{U} \cap \mathcal{C} = \{ \vec{0}_{m \times n}\}$}  \right\}$
 		\State Compute $\vec{B}_{1},\dots,\vec{B}_{(n-\kappa)m+\ell_s-\ell_a}$ a {\it random} $\Fq$--basis of $\left( \mathcal{C}_s \oplus \Amat \right)^{\perp}$ 
 		\State Complete the above basis into a basis $\vec{B}_1,\dots, \vec{B}_{mn-\kappa m + \ell_{s}}$ of $\mathcal{C}_{s}^{\perp}$
 		\State $T \leftarrow \left( \vec{g}, \gamma, \left( \vec{B}_{(n-\kappa)m+\ell_s-\ell_a+1},\dots,\vec{B}_{(n-\kappa)m+\ell_s}\right)\right)$
 		\State\Return $\left( \left( \vec{B}_{1},\dots,\vec{B}_{(n-\kappa)m+\ell_s-\ell_a}\right),T \right)$ 
 	\end{algorithmic}
 \end{algorithm}
\section{Inversion of the Miranda trapdoor~\cite{miranda}} \label{sign}

Here is presented how to sign the hash $\sv\in\Fq^{(n-\kappa)m+\ell_s-\ell_a}$ of a message, that is its associated error $\Ev\in\Fqmn$ with rank at most $\floor{\frac{n-k}{2}}$ with:
$$\left( \tr \left( \vec{E}\vec{B}_{i}^{\top} \right) \right)_{i=1}^{(n-\kappa)m+\ell_s-\ell_a}=\sv$$
where $\vec{B}_{1},\dots,\vec{B}_{m(n-k) +\ell_{s}-\ell_{a}}$ is a basis of a $\mathsf{AddRemove} (\G_\kappa(\gv),\ell_a,\ell_s)$ code. We define $t=\floor{\frac{n-k}{2}}$.

 \begin{itemize}\setlength{\itemsep}{5pt}
 	\item[{\bf Step 1.}] First, we sample uniformly at random $\vec{t} \in \Fq^{\ell_a}$ and we compute by linear algebra a $\vec{Y} \in \Fq^{m \times n}$ such that 
 	$$
 	\left( \tr \left( \vec{Y}\vec{B}_{i}^{\top} \right)
 	\right)_{i=1}^{m(n-\kappa)+\ell_{s}} = (\vec{s},\vec{t})~.
 	$$
 	Notice that we used here a basis $\left(\vec{B}_1,\dots,\vec{B}_{m(n-\kappa) + \ell_{s}}\right)$ of $\mathcal{C}_{s}^{\perp}$ by considering the additional~$\ell_{a}$ matrices from the trapdoor. The point of this step is that if we succeed to decode $\vec{Y}$ in $\mathcal{C}_{s}$ ({\it i.e.,} if we compute $\Ev$ of rank at most $t$ such that~$\vec{Y}-\vec{E} \in \mathcal{C}_{s}$), then we deduce that $	\left( \tr \left( (\vec{Y}-\vec{E})\vec{B}_{i}^{\top} \right)
 	\right)_{i=1}^{m(n-\kappa)+\ell_{s}} = \vec{0}$ and therefore our decoder outputs $\vec{E}$ which is a solution we are looking for. The aim of the next steps is precisely to decode $\vec{Y}$ in $\mathcal{C}_{s}$.

 	\item[{\bf Step 2.}] Given $\vec{Y}$, we compute $\vec{y} \eqdef M_{\gamma}^{-1}(\vec{Y})$ and we try to decode it in the Gabidulin code~$\G_\kappa(\gv)$. If it exists $\Ev$ of rank at most $t$ such that $\vec{Y} - \vec{E} \in \mathcal{C}_{s}$, then we have $\vec{Y} - \vec{E} \in M_{\gamma}\left(\G_\kappa(\gv)\right)$ as $\mathcal{C}_{s} \subseteq M_{\gamma}\left( \G_\kappa(\gv) \right)$. In other words, in such a case we have $\vec{y} - \vec{e} \in \G_\kappa(\gv)$ where~$\vec{e} \eqdef M_{\gamma}^{-1}(\vec{E})$. But~$\vec{e}$ verifies $\norme{\ev} = \norme{\Ev} \leq t$ as $M_{\gamma}$ is an isometry and the decoding algorithm  	of $\G_\kappa(\gv)$ will necessarily return $\vec{e}$ (see Proposition~\ref{propo:decoGab}).

 	\item[{\bf Step 3.}] In the previous step, we decoded $\vec{y}$ in $\G_\kappa(\gv)$. If the decoding algorithm fails, we return to Step~1 and sample another $\vec{t}$. If the decoding algorithm succeeds, we have computed some~$\vec{e}$ with rank at most $t$ such that~$\vec{y} - \vec{e} \in \G_\kappa(\gv)$. Therefore,~$\vec{Y} - \vec{E} \in M_{\gamma}(\G_\kappa(\gv))$ where~$\vec{E} \eqdef M_{\gamma}(\vec{e})$ and we are almost done. To conclude, we just need to verify that~$\vec{Y} - \vec{E} \in \mathcal{C}_{s}$ which may not hold as \emph{a priori} only $\mathcal{C}_{s}\subseteq M_{\gamma}(\G_\kappa(\gv))$ is supposed. Otherwise, we go back to Step~1 and guess another $\vec{t}$. 
 \end{itemize}

The resulting algorithm is the following:

 \begin{algorithm}
 	\caption{$\mathsf{Sign}(\vec{s},T)$: \textsf{Miranda} signing algorithm outputting $\Ev$ of rank at most $t$ such that $\left( \tr\left( \vec{E}\vec{B}_{i}^{\top} \right)\right)_{i=1}^{m(n-\kappa)+ \ell_{s}-\ell_{a}} = \vec{s}$ where the~$\left( \vec{B}_{i} \right)_{i=1}^{m(n-\kappa)+\ell_{s}-\ell_{a}}$ and~$T=\left( \vec{g}, \gamma,(\vec{B}_i)_{i= m(n-\kappa)+\ell_{s}-\ell_{a}+1}^{m(n-\kappa) +\ell_{s}} \right)$ are output by Algorithm~\ref{algo:trap} \label{algo:sgn}}
 	\begin{algorithmic}[1]
 		
 		\State Choose a basis of $\mathcal{C}_s$ such that its first $m(n-\kappa)- \ell_a + \ell_s$ elements are the $\vec{B}_{i}$'s. The remaining ones are denoted $\vec{C}_{1},\dots,\vec{C}_{\ell_a}$
 		\Repeat
 		\State $\vec{t} \getsr \Fq^{\ell_a}$\label{inst:uniformT}
 		\State\label{inst:Y} Compute $\vec{Y} \in \Fq^{m \times n}$: $\left( \tr \left( \vec{Y}\vec{B}_{i}^{\top} \right)
 		\right)_{i=1}^{m(n-\kappa)+\ell_{s}} = (\vec{s},\vec{t}) $
 		\State $\vec{y} \leftarrow M_{\gamma}^{-1}\left( \vec{Y} \right)$
 		\State $\vec{e} \leftarrow \mathsf{Decode}^{\mathsf{Gab}}(\vec{y},\vec{g},\kappa)$ 	\Comment{$\mathsf{Decode}^{\G}$ defined in Proposition~\ref{propo:decoGab}}
 		\State Find $\vec{E} \in \Fq^{m\times n}$ such that  $ \tr\left( \vec{E}\vec{B}_{i}^{\top} \right)_i = \vec{s}$ {\em and} $\tr\left( \vec{E}\vec{C}_{i}^{\top} \right)_i = \vec{t}$
 		\Until $\vec{e} \neq \bot$ \texttt{and} $\left( \tr \left( M_{\gamma}\left( \vec{e}\right)\vec{B}_{i}^{\top} \right)\label{instruct:Until} 
 		\right)_{i=1}^{m(n-\kappa)+\ell_{s}} = (\vec{s},\vec{t}) $
 		\State \Return $M_{\gamma}\left( \vec{e} \right)$
 	\end{algorithmic}
  \end{algorithm}

\end{document}